\documentclass{mn2e}

\input epsf
\begin{document}
\journal{astro-ph/0207213}
\title[Microwave background constraints on inflationary
parameters]{Microwave background constraints on inflationary parameters}
\author[S.~M.~Leach and A.~R.~Liddle]{Samuel M.~Leach$^{1,2}$ and Andrew
R.~Liddle$^1$\\
$^1$Astronomy Centre, University of Sussex, Falmer, Brighton BN1 9QJ, United
Kingdom.\\
$^2$D\'epartment de Physique Th\'eorique, Universit\'e de
Gen\`eve, Quai Ernest-Ansermet 24, CH-1211 Gen\`eve 4,\\ \hspace*{10pt} 
Switzerland (present address).}
\maketitle
\begin{abstract}
We use a compilation of cosmic microwave anisotropy data
(including the recent VSA, CBI and Archeops results), supplemented
with an additional constraint on the expansion rate, to directly
constrain the parameters of slow-roll inflation models. We find
good agreement with other papers concerning the cosmological
parameters, and display constraints on the power spectrum
amplitude from inflation and the first two slow-roll parameters,
finding in particular that $\epsilon_1 < 0.057$. The technique we
use for parametrizing inflationary spectra may become essential
once the data quality improves significantly.
\end{abstract}

\section{Introduction}

Recent measurements of the cosmic microwave background (CMB) show a flat
portion at low multipole number $\ell$ and a sharp peak around $\ell\sim 200$,
as well as
tentative evidence for a peak structure beyond $\ell = 200$. This
represents a tremendous success for the simplest models of the
universe described by a flat Friedmann--Robertson--Walker metric with
adiabatic perturbations, which are in excellent qualitative agreement with these
observations. The power of the CMB is that it can be used
to constrain cosmological parameters, as well as allowing us to test
our assumptions about the form of the initial irregularities in
qualitative and now quantitative ways. The most popular assumption concerning
the initial irregularities is that they originated during a period of
cosmological inflation (see Liddle \& Lyth 2000 for an extensive account of
inflationary cosmology).

There have now been several papers which have searched for possible effects in
this data from quite complicated inflationary models. One example is the
inclusion of extra isocurvature degrees of freedom in the primordial power
spectrum in addition to a dominant adiabatic component (Trotta, Riazuelo \&
Durrer 2001;
Amendola et al.~2002), with the conclusion that the current
data set is consistent with a sub-dominant isocurvature component (or even a
dominant one on large scales in the case where the isocurvature perturbations
are correlated
with the adiabatic ones) and
that the allowed values and ranges of the cosmological parameters are
sensitive to the type of perturbations under consideration. Another example is
attempts to fit inflation-motivated `features in
the power spectrum' to the data (Griffiths, Silk \& Zaroubi 2001;
Barriga et al.~2001; Adams, Cresswell \& Easther 2001). These scalar
power spectra have the
intrinsic property of introducing extra degrees of freedom --- the
shape parameters associated with the feature --- which can be used alter the
peak heights at
will. However it is necessary to choose the features to coincide with
characteristic scales in the CMB power spectrum, such as the first or second
peaks [early work in this direction (Adams, Ross \& Sarkar 1997) was also
motivated by possible features in the matter power spectrum]. The CMB
spectrum alone offers no evidence for any extra features and so smooth power
spectra are currently best motivated.

It is surprising that relatively little attention has been paid to the CMB
spectrum
resulting from slow-roll inflation, even though these models have been the most
intensively studied since its conception. Slow-roll inflation has
acted as a guiding principle for inflation model builders, and is the simplest
assumption and capable of giving excellent agreement with observations. The
reason why specific studies of slow-roll inflation have been lacking is the
usual assumption that inflation predicts a nearly power-law shaped
spectrum, and hence that any information about inflation can be extracted
as some linear combination of the constraints on the two key
parameters, the scalar spectral index $n_{{\rm S}}$, and the tensor
fraction $R$, an approach used by Kinney, Melchiorri \& Riotto (2001) and by
Hannestad et al.~(2002) to discuss
constraints on inflation. Hansen \& Kunz (2001) also included the running of the
spectral index, translating constraints on these parameters to place bounds on
derivatives of the inflaton potential.
Other parameter analyses (Wang, Tegmark \& Zaldarriaga
2002; Percival et al.~2002) have tended to focus on results for other
cosmological parameters
such as the densities of the various matter components, and have been content to
use this simple parametrization.

However, this approach ignores the fact that current data are only weakly
constraining, and the current data set permits parameter regions where a
significant deviation from a
Harrison--Zel'dovich spectrum is allowed, and where the use of the full
slow-roll power spectra is required to obtain robust results. The principal aim
of this paper is to make the first direct estimation of slow-roll inflation
parameters from CMB data.
While the full predictions are relevant only in extreme regions of parameter
space given current data, as the global cosmological data set improves (and in
particular as significant observational weight develops on the
high $\ell$ part of the CMB spectrum) it may well be that these types of
corrections take on increasing importance, depending which (if any) inflation
models prove capable of fitting the data.

We do not expect any dramatic new results from this analysis given the quality
of present data, though it is a useful test of the robustness of results under
more general forms of the initial power spectra. However it is an important test
of principle to bring these methods to bear on cosmological parameter
estimation, as the high-quality data of coming years may well require the
high-accuracy description of the power spectrum that this approach allows.

\section{Inflationary parameters}\label{inf_params}

In order to implement the inflationary cosmology into a parameter
search method, we need an adequate parametrization of the scalar and
tensor power spectra that give rise to the observed anisotropies. We use the
recently introduced horizon-flow parameters
(Schwarz,
Terrero-Escalante \& Garcia 2001) which are based around the Hubble
parameter during inflation and its derivatives. These parameters enter
directly into both the Friedmann equation and into the mode equations
for scalar  and tensor perturbations. They often simplify the results
from analytical calculations of the power spectrum because of their
simple definition
\begin{eqnarray}
\epsilon_0 & = & H_{{\rm inf}}/H \,;\\
\epsilon_{i+1} & \equiv & \frac{d\ln |\epsilon_i|}{dN}\;,\; i \geq 0
\end{eqnarray}
where $H_{{\rm inf}}$ is the Hubble parameter at some chosen time and $N$ is the
number of $e$-foldings of inflation.
More importantly from the observational side, this formalism provides
a consistent and detailed description of the shapes of the scalar and
tensor power spectra as well as their absolute and relative
normalizations, independent of other cosmological parameters (such as the
cosmological constant physical density $\omega_{\Lambda}$).

Our inflationary parameter set consists of the parameters
$\{H_{{\rm inf}}$, $\epsilon_1$, $\epsilon_2$,
$\epsilon_3,\ldots\}$, evaluated at some particular time during inflation. As
the amplitude of inflationary perturbations is primarily determined around
horizon crossing, we can relate time to scale by asking when a given scale
equalled the Hubble radius during inflation, $k=aH$, and we can then think of
these parameters as a function of scale. The scale at which they will be
evaluated is in a sense arbitrary but is of course most wisely chosen to be
around the centre of the scales actually probed by the observations.
This parameter set replaces that made up of astrophysical quantities
$\{{\cal P}_{\cal R}$, $R_{10}(\Omega_{\Lambda},h)$, $n_{{\rm
S}}-1$, $dn_{{\rm S}}/d\ln k$, $n_{{\rm T}}$, $dn_{{\rm T}}/d\ln k,\ldots\}$
that are used if the power spectra are taken as the starting point.
Here ${\cal P}_{\cal R}$ is the amplitude of scalar perturbations specified at
the scale $k = 0.05 \, {\rm Mpc}^{-1}$, $R_{10}$
is the ratio of the tensor and scalar $C_{\ell}$ curves evaluated at
$l=10$, and $n_{{\rm S}}-1$ and $n_{{\rm T}}$ are the slopes
of the initial scalar and tensor power-law spectra.
Use of the inflationary parameters automatically enforces the
inflationary consistency relations between scalar and tensor
perturbations, and so there are fewer parameters to be fit; a full
treatment of observations would also test these conditions though
present data is not good enough for a meaningful confirmation.

The current data set has very limited abilities to constrain tensor
perturbations and usually only $R$ is used to describe them (perhaps coupled
with a consistency relation to fix the tensor spectral index). It is also
common, though not universal, for parameter searches to neglect scale-dependence
of the scalar spectral index.

The power spectrum ${\cal P}$ from slow-roll inflation can be obtained as an
expansion of the power spectrum (or some function of the power spectrum)
in terms of the logarithmic wavenumber. While the usual power-law expression is
mostly simply related to an expansion of $\ln {\cal P}$, since it is the power
spectrum itself which is constrained by observations it is most direct to expand
${\cal P}(k)$ itself (Martin, Riazuelo \& Schwarz 2000; Leach et al.~2002).
The spectra of curvature perturbations ${\cal P}_{{\cal R}}$ and of tensor
perturbations ${\cal P}_h$ are written as
\begin{equation}
\label{eqn:chaotic}
{\cal P}_{{\cal R},h} \propto b_0 + b_1 \ln \left(k\over k_*\right)
         + \frac{b_2}{2} \ln^2\left(k\over k_*\right)
         + \dots,
\end{equation}
The detailed predictions for the $b_i$ have been calculated to second
order in the slow-roll parameters (Stewart \& Gong 2001; Leach et al.~2002) to
be
\begin{eqnarray}
b_{{\rm S}0} &=&  \label{bs0}
        1 - 2\left(C + 1\right)\epsilon_1 - C \epsilon_2
        + \left(2C^2 + 2C + {\textstyle\frac{\pi^2}{2}} - 5\right)
        \epsilon_1^2 \nonumber \\
    & & + \left(C^2 - C + {\textstyle\frac{7\pi^2}{12}} - 7\right)
        \epsilon_1\epsilon_2 
        + \left({\textstyle\frac 12} C^2 + {\textstyle\frac{\pi^2}{8}}
        - 1\right)\epsilon_2^2 \nonumber \\
    & & + \left(-{\textstyle\frac 12} C^2
        + {\textstyle\frac{\pi^2}{24}}\right)\epsilon_2\epsilon_3 \\
b_{{\rm S}1} &=&
        - 2\epsilon_1 - \epsilon_2 + 2(2C+1)\epsilon_1^2
        + (2C - 1)\epsilon_1\epsilon_2 \nonumber \\
    & & + C\epsilon_2^2 - C\epsilon_2\epsilon_3 \\
b_{{\rm S}2} &=& 4\epsilon_1^2 + 2\epsilon_1\epsilon_2 + \epsilon_2^2 -
    \epsilon_2\epsilon_3
\end{eqnarray}
for the scalars, and
\begin{eqnarray}
b_{{\rm T}0} &=& 1 - 2\left(C + 1\right)\epsilon_1
        + \left(2C^2 + 2C + {\textstyle\frac{\pi^2}{2}} - 5\right)
        \epsilon_1^2 \nonumber \\
    & & + \left(-C^2 - 2C + {\textstyle\frac{\pi^2}{12}} - 2\right)
        \epsilon_1\epsilon_2 \\
b_{{\rm T}1} &=&
        - 2\epsilon_1 + 2(2C + 1)\epsilon_1^2
        - 2(C + 1)\epsilon_1\epsilon_2 \\
b_{{\rm T}2} &=& 4\epsilon_1^2 - 2\epsilon_1\epsilon_2 \label{bt2}
\end{eqnarray}
for the tensors, where $C \equiv \gamma_{\rm E} + \ln 2 - 2 \approx -0.7296$ is
a numerical constant. The full parametrization of
inflation as given here, where we truncate at $\epsilon_3$, should remain
sufficiently accurate for some time to come, quite likely including {\it Planck}
satellite results.

Concerning the relative normalization of tensor and scalars, the
constant of proportionality for Eq.~(\ref{eqn:chaotic}) is given by
the slow-roll amplitudes
\begin{eqnarray}
\label{eqn:pnorm}
{\cal P}_{h0}(k_*) &=& {16 H_{{\rm inf}}^2\over \pi m_{\rm Pl}^2},\\
{\cal P}_{{\cal R}0}(k_*) &=& {H_{{\rm inf}}^2 \over \pi \epsilon_1
m_{\rm Pl}^2}\,.
 \label{eqn:psnorm}
\end{eqnarray}
In the slow-roll
limit the tensor to scalar ratio is given by ${\cal P}_{{\rm h}}/{\cal
P}_{{\cal R}} = 16\epsilon_1$, and the more general expression can be
read from the above expansion  coefficients as
\begin{equation}
\label{eqn:R2}
\frac{{\cal P}_{{\rm h}}}{{\cal P}_{{\cal R}}} =
16\epsilon_1[1+C\epsilon_2+\dots] \,.
\end{equation}

This logarithmic expansion of the power spectra is accurate over a
wide range of the inflationary parameter space. However, unlike
the expansion of $\ln{\mathcal P}(k)$ (the power-law expansion),
the logarithmic expansion of ${\mathcal P}(k)$ can become negative
at large $|\ln(k/k_*)|$ for large $\epsilon_{i}$. This
pathological behaviour serves as a warning against using either
expansion alone for robustly extracting any inflationary signal in
this regime without cross-checks. In any case, this is behaviour
does not occur in our analysis due to the limited range of scales
currently probed by the CMB. The methods that we describe in this
paper therefore complement the traditional approach, and a careful
analysis should utilize both (Leach et al.~2002).

To lowest order in slow-roll these predictions reduce to the well-known
linearized form
\begin{eqnarray}
n_{{\rm S}}-1 &=& -2\epsilon_1-\epsilon_2,\label{lin1}\\
n_{{\rm T}} &=& -2\epsilon_1,\\
R\equiv\frac{{\cal P}_{{\rm h}}}{{\cal P}_{{\cal R}}} &=& 16\epsilon_1.
\label{lin3}
\end{eqnarray}
Written is this way it is clear that an observational strategy
would ideally use information from the tensor sector (primarily the tensor
amplitude) to break the degeneracy between $\epsilon_1$ and
$\epsilon_2$ for the scalar spectral index. The effect of
increasing $\epsilon_1$ is to boost the large-angle anisotropies via the tensor
amplitude while
simultaneously tilting downwards the small-angle anisotropies via the scalar
tilt,
which act roughly constructively.

In slow-roll inflation it is possible to relate
the horizon-flow parameters to the shape of the inflationary potential
\begin{eqnarray}
H^2 &\simeq& \frac{8\pi}{3 m_{\rm Pl}^2} V, \\
\epsilon_1 &\simeq& \frac{m_{\rm Pl}^2}{16 \pi} \left({V'\over V}\right)^2,\\
\epsilon_2 &\simeq& \frac{m_{\rm Pl}^2}{4 \pi}
 \left[\left({V'\over V}\right)^2 - {V''\over V}\right], \\
\epsilon_2 \epsilon_3 &\simeq& \frac{m_{\rm Pl}^4}{32 \pi^2} \left[
 {V''' V'\over V^2} - 3{V''\over V}\left({V'\over V}\right)^2\!
 + 2 \left({V'\over V}\right)^4\right]\! . \ \ \ \
\end{eqnarray}

\section{Observational constraints}

We fit to a CMB dataset comprising data from COBE, BOOMERanG,
Maxima, DASI, VSA, CBI and Archeops. We follow the method of
Lesgourgues \& Liddle (2001) in defining a simple $\chi^2$ with
penalty functions over $D_{{\rm \ell}}\equiv (\Delta T_{{\rm
\ell}})^2= \ell(\ell+1)C_{{\rm \ell}}/2\pi$ for the COBE (Bennett
et al.~1996; Tegmark \& Hamilton 1997), BOOMERanG (Netterfield et
al.~2002), Maxima (Lee et al.~2001) and Archeops (Benoit et
al.~2002) data of the form
\begin{eqnarray}
\chi^2= \sum_\ell\frac{\left(D^{{\rm theo}}_{{\rm \ell}}-
   (1+c\sigma_{{\rm c}}+b\sigma_{{\rm b,l}})D^{{\rm obs}}_{{\rm \ell}}\right)^2}
    {\sigma^2_{{\rm l}}}+b^2+c^2,
\end{eqnarray}
where for COBE we have $\sigma_{{\rm b,l}}=\sigma_{{\rm c}}=0$,
BOOMERanG $\sigma_{{\rm c}}=0.20$ and $\sigma_{{\rm
b,l}}=0.43\times 10^{{-6}}\ell^2$, Maxima $\sigma_{{\rm
b,l}}=2\times10^{{-6}}\ell^{1.7}$ and $\sigma_{{\rm c}}=0.08$, and
Archeops $\sigma_{{\rm c}}=0.14$ and $\sigma_{{\rm b,l}}=0$. For
the DASI (Halverson et al.~2002), VSA (Scott et al.~2002) and CBI
(Pearson et al.~2002) (`mosaic' configuration, odd binning up to
$\ell=1500$) data we define the $\chi^2$ to be
\begin{eqnarray}
\chi^2=& \left(D^{{\rm theo}}_{{\rm i}}-
    (1+c\sigma_{{\rm c}})D^{{\rm obs}}_{{\rm i}}\right)
    M^{-1}_{{\rm ij}} \nonumber \\
    &\left(D^{{\rm theo}}_{{    \rm j}}-
    (1+c\sigma_{{\rm c}})D^{{\rm obs}}_{{\rm j}}\right)
    + c^2.
\end{eqnarray}
where $\sigma_{{\rm c}}=0.08, 0.06, 0.10$ respectively and the
correlation matrix is given by
\begin{equation}
M_{{\rm ij}}=\Delta D_{{\rm i}}V_{{\rm ij}}\Delta D_{{\rm j}}+
      \sigma^2_{{\rm c}}D_{{\rm i}}D_{{\rm j}}.
\end{equation}
We assume gaussian window functions, except for the CBI data for
which the published window functions are used, and the Archeops
data for which the window functions are taken to be top-hat
functions. All error bars are approximated as symmetric, taking the
larger error bar when they are not. We analytically determine the
coefficients $b_{{\rm boom}}$, $c_{{\rm boom}}$, $b_{{\rm max}}$,
$c_{{\rm max}}$, $c_{{\rm dasi}}$ (simply by simultaneously
solving equations such as $\partial\chi^2/\partial b_{{\rm
boom}}=0$) and then sum the $\chi^2$ over all the experiments.

The CMB models were calculated using the January 2002 version of CAMB
(Lewis, Challinor \& Lasenby 2000) which can be easily be modified to
incorporate
absolute and relative normalizations of the scalar and tensor power
spectra.\footnote{A module
to directly input the predictions of slow-roll inflation to the {\sc camb}
program is available to download at {\tt
www.astronomy.sussex.ac.uk/$\sim$sleach/inflation/}}
We note in passing that indeed the correct amplitude
in the tensors is more important that the correct tensor tilt. Even when we move
the pivot scale away from COBE scales where the tensor spectrum is
physically relevant, the tensor to scalar ratio only runs weakly back
to the COBE scales
\begin{equation}
\frac{{\cal P}_{{\rm h}}}{{\cal P}_{{\cal R}}}=16\epsilon_1[1+
(\ln(k/k_*)+C)\epsilon_2+\dots].
\end{equation}
The actual shape of the tensor spectrum will have very little effect
at this stage, and so we are certainly justified in taking the same
pivot scale for both the scalar and tensor spectra.

We examine a set of spatially-flat cosmologies using the
parameters $\{\omega_{{\rm B}},\omega_{{\rm M}}, \omega_{\Lambda},
e^{-2\tau}, {\mathcal P}_{{\mathcal R}},\epsilon_1,\epsilon_2\}$
where $\omega_{{\rm i}}\equiv \Omega_{{\rm i}}h^2$ measures the
physical matter density and $\tau$ is the optical depth to the
last-scattering surface. Our grid runs overs the range
$0.006<\omega_{{\rm B}}<0.045$, $0.03<\omega_{{\rm M}}<0.28$,
$0.0<\omega_{\Lambda}<1.2$, $0.28<e^{-2\tau}<1.0$, $7<{\mathcal
P}_{{\mathcal R}}\times 10^{10}<48$, $0.0001<\epsilon_1<0.07$,
$-0.4<\epsilon_2<0.3$, with a uniform spacing of
$8^3\times7\times22\times8\times11$. It is worth emphasising that
we have included ${\mathcal P}_{\mathcal R}$ as a full parameter
instead of using an analytic procedure to choose the best
amplitude from the data and each model. Apart from our obvious
interest in primordial physics, ${\mathcal P}_{\mathcal
R}e^{-2\tau}$ is a parameter that will be determined with high
accuracy in the future and so should be included in the analysis.

The dependence of the power spectrum on the third slow-roll
parameter $\epsilon_3$ is very weak, given the current data set,
and so we fix $\epsilon_3=0$ while still using the second-order
expressions, which give a better representation of the power
spectrum at large $\epsilon_1,\epsilon_2$ and $|\ln(k/k_*)|$ than
the first-order expressions. Our decision to truncate the
slow-roll expansion after $\epsilon_2$ is based on subtle
considerations. As the series expansion is an infinite one, it
clearly has to be truncated somewhere; if we did include
$\epsilon_3$ the same question of whether to include or not would then arise for 
$\epsilon_4$.
We believe that a suitable criterion is that one ought not to
include parameters which the data cannot distinguish from zero,
i.e.~we impose a null hypothesis that those parameters are zero.
While that criterion is driven by data alone, it makes sense in
the context of inflation because one does expect the parameters to
be small, and if the data is unable to constrain them to be small,
then the parameter space explored would include models which would
not be satisfactory inflation models. Unfortunately this criterion
cannot be rigidly applied, as even $\epsilon_1$ and $\epsilon_2$
cannot presently be distinguished from zero; thirty years of
large-scale structure studies have not excluded the
Harrison--Zel'dovich spectrum. However if inflation is to be
meaningfully tested, eventually at least one of these parameters
needs to be detected with a non-zero value and so we do include
them in our current treatment. This is appropriate as inflation
models do predict values of $\epsilon_1$ and $\epsilon_2$
comparable to constraints from present data. By contrast, typical
inflation models predict $\epsilon_3$ to be much smaller than the
present limits from data (see e.g.~Hansen \& Kunz 2001). Whether
experiments such as {\it Planck} can detect a non-zero
$\epsilon_3$ remains to be seen.

We fix the pivot point to $k_*=0.05$Mpc$^{-1}$, ignoring the
cosmological dependence, proportional to $\omega_{{\rm M}}^{1/2}$,
of the range of scales probed by the CMB. The Hubble parameter is
an auxiliary parameter given by
\begin{equation}
h=\sqrt{\omega_{{\rm M}}+\omega_{{\Lambda}}}
\end{equation}
Throughout this paper we examine the effect of applying the HST Hubble
parameter prior of $h=0.72\pm0.08$ at 1$\sigma$, which acts as a
constraint on the total physical matter density for our flat cosmologies.

\begin{figure}
\centering
\leavevmode \epsfysize=18cm \epsfbox{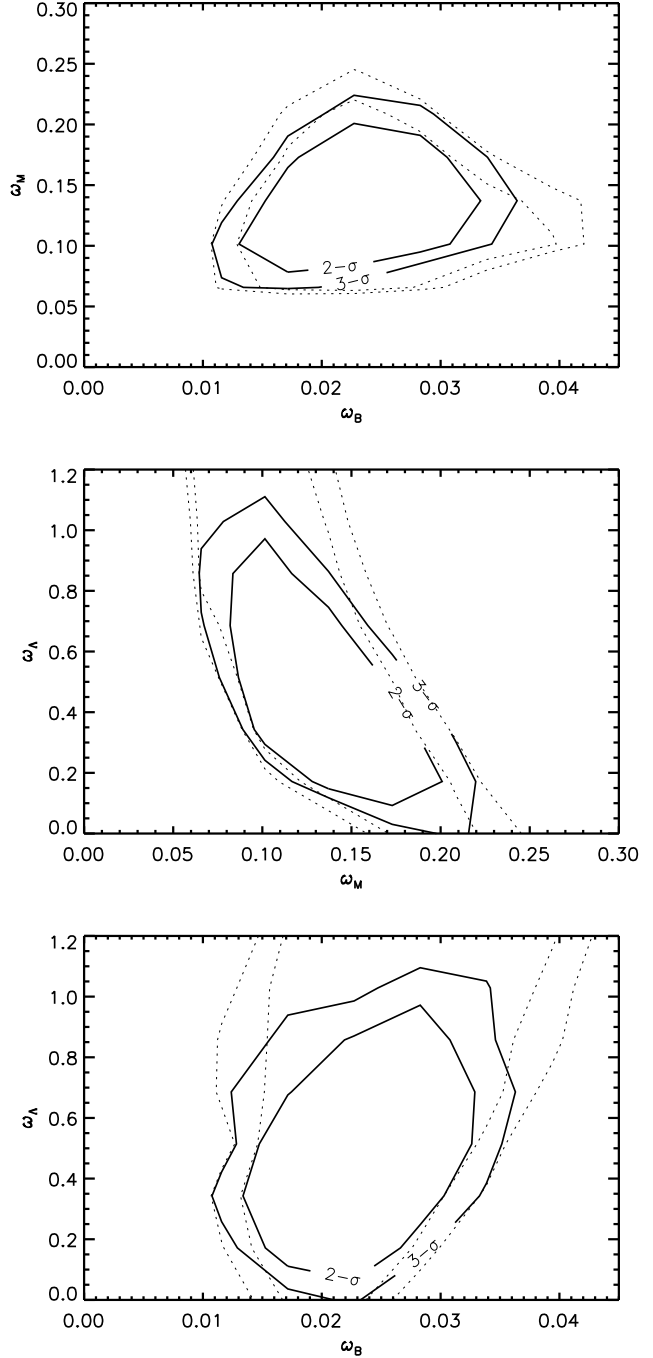}\\
\caption[obh2_omh2]{\label{f:cosm} Three views of the constraints on
$\omega_{{\rm B}}$, $\omega_{\rm M}$ and $\omega_{\Lambda}$. The
dashed lines show the constraints from the CMB alone, while the solid
lines include the HST prior on the Hubble constant.}
\end{figure}

When presenting parameter constraints, we simply minimize the
$\chi^2$ over unwanted parameters. In principle one should
consider performing a cubic spline over the $\chi^2$ of our coarse
grid of models in order to examine the shape of the likelihood
function $e^{-\chi^2/2}$ and to subsequently perform a
marginalization procedure (Efstathiou et al.~1999). However,
minimizing the $\chi^2$ reproduces the basic shape of the allowed
region in parameter space (Tegmark \& Zaldarriaga 2000). Our
best-fit model has $\chi^2=51.1$ and so we plot the 2$\sigma$ and
3$\sigma$ contours for a $\chi^2$ distribution with 51.1 degrees
of freedom. We omit the 1$\sigma$ contour, as little useful
information is conveyed by this level of certainty in the context
of CMB parameter searches. We regard the models enclosed by the
2$\sigma$ contours as representing good fits to the data, and the
3$\sigma$ contours mark out the region where tension between the
data and the models begins.

Turning firstly to three of the standard cosmological parameters,
the densities of the three main components, we see in
Fig.~\ref{f:cosm} impressive agreement with the standard BBN value
of $\omega_{{\rm B}}=0.020\pm0.002$ at 95 per cent confidence
(Burles, Nollett \& Turner 2001). This acts as a useful
consistency check on the assumption of adiabaticity, given that
the inclusion of a subdominant isocurvature mode tends to widen
the allowed range in $\omega_{{\rm B}}$ (Trotta et al.~2001).
For the  current dataset we observe a weak correlation between the
constraints on $\omega_{{\rm M}}$ and $\omega_{\Lambda}$. The main
effect of applying a prior on the Hubble parameter, $h$, is to
rule out models with a large physical density in
$\omega_{\Lambda}$, and to close the contours near
$\omega_{\Lambda}=0$. When quoting parameter constraints we
include the HST prior.

\begin{figure}
\centering
\leavevmode\epsfysize=5.8cm \epsfbox{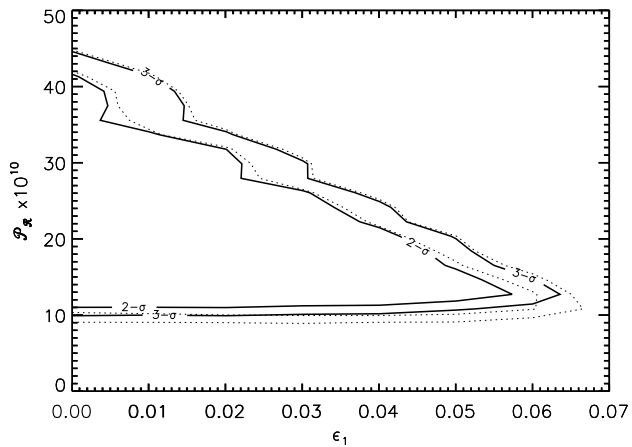}\\
\caption[Pseps1]{\label{Pseps1}Constraints on the amplitude ${\mathcal
P}_{{\mathcal
R}}$ and the first slow-roll parameter $\epsilon_1$, the solid lines again 
including the HST prior.}
\end{figure}

We now turn to the inflationary parameters, which are the main focus of our
study. In Fig.~\ref{Pseps1} we
plot the constraints on ${\mathcal P}_{{\mathcal R}}$ and
$\epsilon_1$. The general trend is that as we increase the tensor
component, $R=16\epsilon_1$, which contributes at low $\ell$, then
we must decrease the scalar normalization in order to continue to fit the low
$\ell$ data. Including ${\mathcal P}_{{\mathcal R}}$ as a full
parameter in the analysis
acts as a
useful warning against including higher-order parameters in the power
spectrum; until we have pinned down the amplitude of perturbations at
some given scale, there is little sense in trying to measure the
higher derivatives of the power spectrum such as the running of the
spectral index etc, under the assumption that these higher derivatives
are weak. Reading off the 2-$\sigma$ bounds on ${\mathcal
P}_{{\mathcal R}}$ and $\epsilon_1$ we find
\begin{eqnarray}
&& 11< {\mathcal P}_{{\mathcal R}} \times 10^{10}   <42,\\
&& \epsilon_1 <0.057 \,.
\end{eqnarray}
These are the main results of this paper, and are in good agreement with the
analysis of Lewis \& Bridle (2002). The upper bound on $\epsilon_1$ is
consistent with the
inflationary hypothesis, which requires $\epsilon_1<1$ for inflation
to occur.
It isn't possible push this limit much further down using
CMB data from ground and balloon observations due to the
calibration uncertainties in the high $\ell$
region which mask the primary effect of $\epsilon_1$, which is to boost
the low $C_\ell$s relative to the high $C_\ell$s.

Reading off the values of ${\mathcal P}_{\mathcal R}$ and $\epsilon_1$
at the tip of the $2$-$\sigma$ contours of Fig.~\ref{Pseps1} allows us to
place a constraint on the energy scale of inflation
\begin{eqnarray}
\frac{H_{{\rm inf}}}{m_{{\rm Pl}}} &<& 1.5 \times 10^{-5},\\  
V_{{\rm inf}}^{1/4}&<& 2.9\times10^{16} \, \textnormal{GeV}.
\end{eqnarray}
Knox \& Song (2002) and Kesden, Cooray \& Kamionkowski (2002) have
calculated a limit on $V_{{\rm inf}}^{1/4}$ below which tensors
can not be detected directly in the $B$-mode polarization of the
CMB due to a contaminating signal from lensing of the $E$-mode
along the line of sight. Combining this with our upper limit then
we find that the inflationary energy scale must lie in the range
\begin{equation}
3<\frac{V_{{\rm inf}}^{1/4}}{10^{15}\textnormal{GeV}}< 29
\end{equation}
in order to directly detect the tensor spectrum via the $B$-mode
polarization of the CMB.

\begin{figure}
\centering
\leavevmode\epsfysize=5.8cm \epsfbox{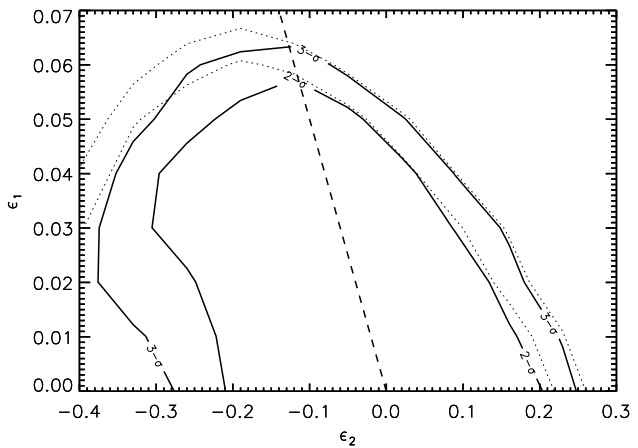}\\
\caption[e1e2]{\label{e1e2} Constraints on $\epsilon_1$ and
$\epsilon_2$. The data are consistent with a scale-invariant,
tensorless spectrum ($\epsilon_1<<1$, $|\epsilon_2|<<1$), but a
significant tensor fraction and tilt are still permitted by the
data. The dashed line, $\epsilon_1 = - \epsilon_2/2$, indicates
inflationary models with a scale-invariant power spectrum. To the
right the spectra are red, to the left they are blue.}
\end{figure}

In Fig.~\ref{e1e2} we plot the constraints on $\epsilon_1$ and
$\epsilon_2$, the first two slow-roll parameters, showing that a
wide range of slow-roll inflation models fit the present data. The
data are consistent with a scale-invariant scalar spectrum with no
tensors ($\epsilon_1\ll 1$, $|\epsilon_2| \ll 1$). We can also
read off an approximate and loose bound on the second slow-roll
parameter
\begin{equation}
-0.31<\epsilon_2<0.2 \,.
\end{equation}
To guide the eye we plot the line $\epsilon_1=-\epsilon_2/2$ along
which the inflationary scalar power spectrum is approximately
scale-invariant. As can be expected, the contours lean in the same
direction as this line reflecting the main inflationary degeneracy
of Eq.~(\ref{lin1}).

\begin{figure}
\centering
\leavevmode\epsfysize=5.8cm \epsfbox{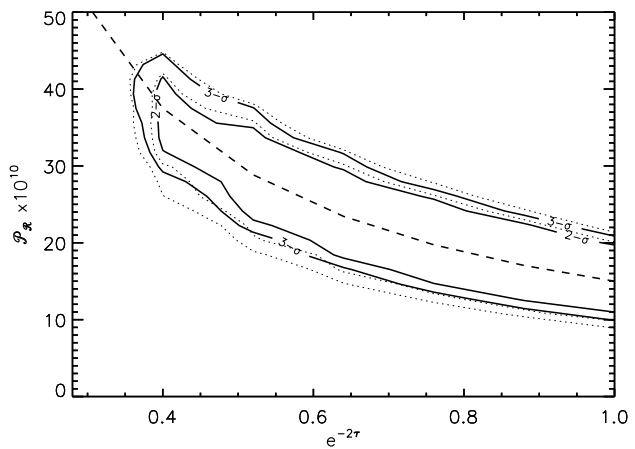}\\
\caption[Pstau]{\label{Pstau} Constraints on ${\mathcal
P}_{{\mathcal R}}$ and the optical depth $\tau$. The thick dashed
curve is given by $10^{10}{\mathcal P}_{{\mathcal
R}}e^{-2\tau}=15$.}
\end{figure}

Finally, in Fig.~\ref{Pstau} we plot the constraints on ${\mathcal
P}_{{\mathcal R}}$ and optical depth $\tau$ revealing the
anticipated degeneracy between these two parameters, both of which
affect the amplitude of the acoustic peaks. We derive the limit
$0.4<e^{-2\tau}(<1.0)$ corresponding to an upper limit
$\tau<0.45$.

\section{Summary}

We have implemented the detailed second-order predictions for the
inflationary power spectra, given by equations (\ref{bs0}) to
(\ref{eqn:psnorm}), into a CMB parameter search method, using the
logarithmic expansion of ${\mathcal P}(k)$,
Eq.~(\ref{eqn:chaotic}), for the first time. Although the present
data set cannot hope to actually measure the weak running of the
spectral index induced by a significant tensor component, we
derived sensible limits on the power spectrum amplitude ${\mathcal
P}_{{\mathcal R}}(k=0.05\,{\rm Mpc}^{-1})$ and the first two
parameters, $\epsilon_1$ and $\epsilon_2$, by assuming the running
to be weak, which was achieved by considering models with
$\epsilon_1<0.07$, $-0.4<\epsilon_2<0.3$ (and fixing
$\epsilon_3=0$). We also derived a sensible limit for
$\omega_{{\rm b}}$ which acts as a useful consistency check on the
assumption of adiabatic perturbations with an approximately
power-law form.

While the results of the present paper do not add much to existing
studies parametrizing the spectra as power-laws, our paper
represents an important point of principle in implementing precise
slow-roll inflation predictions for the first time. As the global
dataset improves, including MAP and then {\it Planck} data, it is
quite likely that these techniques are required to ensure robust
estimation even of cosmological parameters such as the densities
of the various components. Further, these techniques will be
essential to squeeze the maximum possible amount of information
out of the data regarding inflation, should slow-roll inflation
continue to give the simplest viable interpretation of
observational data. As the data-set improves, it will be
interesting to open up the $\epsilon_3$ direction as well as
exploring the possibility of a negligible tensor prior, in order
to differentiate between inflationary models as effectively as
possible. An interesting goal would be to determine the signs of
both $\epsilon_2$ and $\epsilon_3$, which would allow us to
immediately rule out three quarters of all single-field slow-roll
inflation models.

\section*{ACKNOWLEDGMENTS}

SML was supported by PPARC and ARL in part by the Leverhulme Trust. We thank
Louise Griffiths, Julien Lesgourgues, Karim Malik, J\'er\^ome Martin and Dominik
Schwarz for discussions.



\bsp
\end{document}